# Some results and open problems in research on low dimensional organic conductors

V.CELEBONOVIC
*Institute of Physics, Pregrevica 118,11080 Zemun-Beograd, Yugoslavia*
*e-mail:* vladan@phy.bg.ac.yu

This paper broadly reviews some of the fundamental results of research on a family of quasi one-dimensional organic conductors called the Bechgaard salts. To a lesser extent, the quasi two-dimensional salts will also be discussed. Apart from a selection of known results, drawn from the literature and the author's research experience, some open problems will also be presented.

## 1. Introduction

This contribution is a review of some of the basic results of research on the low dimensional organic conductors. As this field of research is huge, the choice of the topics to be covered will be conditioned by scientific interest and the author's research experience.

In these materials, the electrical conductivity is directed mainly along one axis, (quasi one-dimensional, Q1D) or within a plane (quasi two-dimensional, Q2D). The "quasi" appears in the names of these materials because the conductivity along other directions, apart from the axis and the plane mentioned above, is NOT strictly equal to zero.

Some brief historical notes are in order at the beginning. The first to pose the question about the possible existence of superconductivity in organic materials was London [2], but the field remained mostly speculative until the synthesis of the perylene bromine complex [3], which had a high conductivity compared to other materials of the time. A big "boost" of this field occurred in 1964, with Little's model of a possible high temperature conductor [4]. He theoretically "devised" a material consisting of an organic polymer with highly polarizable side chains. Electrons could move along these chains with a phonon-mediated attractive interaction. He showed that such a material would become superconducting under certain conditions, and that the transition temperature could be raised arbitrarily by a change of parameters. Such a system was never synthetized, and the main achievement of this model was to direct attention towards Q1D systems.

A high electrical conductivity was discovered at the beginning of the seventies in the molecule TTF-TCNQ, (tetrathiafulvalene tetracyanoquinodimethane, or explicitely $C_{18} H_8 S_4 N_4$) [5] . Not only was the conductivity high, but the material was Q1D. For this particular material, the conductivity increases dramatically near $T \sim 60K$ but it then undergoes a phase transition into an insulating state.

## 2. The Bechgaard salts

The general formula of the Bechgaard salts is $(TMTSeF)_2X$, where $(TMTSeF)_2$ denotes bis-tetramethyl- tetraselenafulvalene and $X$ is an anion such as $X=PF_6$, $FSO_3, ReO_4, ClO_4$...[6]. The first organic superconductor [6] was discovered for $X=PF_6$. It was soon shown that in this material the transition to the superconducting state occurrs at $P = 1.2$ GPa, and $T = 0.9$ K.

Already early experiments have shown that the electrical conductivity of the Bechgaard salts is extremely anisotropic [1]. The axis (called the a-axis) along which the conductivity is greatest corresponds to the direction in which the overlap of the $\pi$ orbitals is strongest, while the coupling is the weakest in the direction along which the anions and methyl groups separate the main $\pi$ orbitals of the TMTSeF molecules. The ratio of conductivities $\sigma_a:\sigma_c$ along these two axes can be as high as 1000:1.

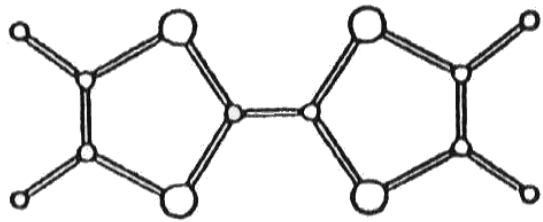

*Figure 1.* The structure of TMTSeF

The structure of the TMTSeF molecule is shown on fig.1, where the bigger circles denote Se, and the smaller C atoms. This family of salts was later expanded [1]. The "recipes" for their synthesis are complex, but can be found in the literature (for example [7]). There also exist some "fine details".

For example, the personal research experience of the author shows that specimens synthetized in Copenagen and Orsay are not of the same quality. The main result which emerged from early work and was amply confirmed later was that the electrical resistivity of the Bechgaard salts is clearly non-ohmic, and accordingly can not be theoretically described by the standard theory of metals.

Experiments are usually done by the "four probe" method. Contacts on a specimen are made by inserting specimen enclosed in special masks in gold vapour. On these contacts are glued gold wires whose diameter is only around 30 microns. Note that specimen are considered as "big" if their dimensions are of the order of 0.5 x 0.5 x 5 mm. This brief description illustrates the complexity of the preparation of the experiments. The experiments are simple in principle, but in fact complicated. However, they give the possibility for determining the electrical conductivity of the Bechgaard salts.

Following the changes of conductivity as a function of various external parameters (such as the pressure, temperature or magnetic field ) gives the possibility to deduce phase transition points and lines on the phase diagram of the Bechgaard salts. An example of the experimental results ,taken from the historic discovery paper [6] is provided in fig. 2.

Another phenomenon noted in studies of the Bechgaard salts is anion ordering. It occurs in salts whose anions have tetrahedral symmetry. These anions can take two orientations with respect to the crystal lattice. Experiments on salts with anions $FSO_3$ and $ClO_4$ have shown that the anion ordering temperature in these two materials increases with the applied pressure [8,9], although simple physical logic would lead to contrary expectations.

Data for the salt $(TMTSeF)_2FSO_3$, taken from experimental data accumulated for [8] are shown in fig. 3. The solid line is a fit of the data to a fourth degree polynominal. To the best of the author's knowledge no plausible physical explanation of this behaviour has been proposed.

Some years ago [10], the thermal conductivity of the Bechgaard salts was measured for the first time. Measurements of thermal conductivity are a useful method in determining the gap structure in various unconventional superconductors This experiment was performed on $(TMTSeF)_2ClO_4$ in the superconducting, metallic and insulating phases. It was found that the electronic contribution to heat transport decreases below the critical temperature. The experimental results were interpreted as indicating that the superconducting gap function has no nodes.

## 3. Theoretical studies

Soon after their discovery, it became apparent that the Bechgaard salts were theoretically "non standard" materials. A step forward was the application of the Hubbard model. The starting Hamiltonian of this model in one dimension has the form:

$$H = -t\sum_{l,\sigma}(c^+_{l+1,\sigma}c_{l\sigma} + .*) + U\sum_{l}n_{l\uparrow}n_{l\downarrow} \quad (1)$$

where the sign '*' denotes complex-conjugate terms, $t$ is the electron transfer energy (the so called hopping) and $U$ is the energy of interaction of electrons with opposite spins on different ions.

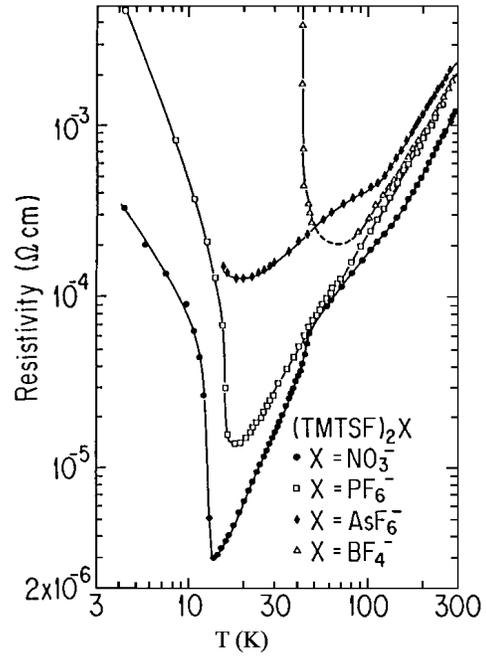

*Figure 2.* The temperature dependence of the d.c. resistivity of several Bechgaard salts [6].

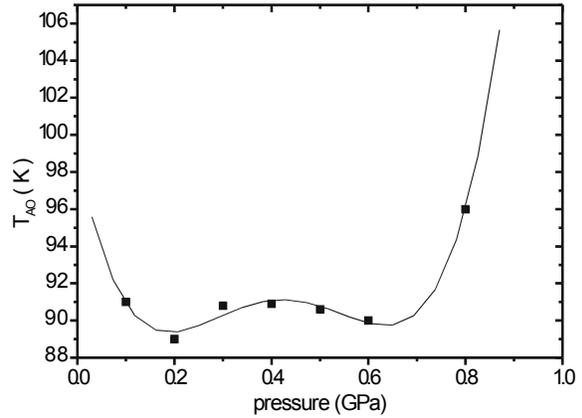

*Figure 3.* Anion ordering temperatures under increasing pressure for $(TMTSF)_2FSO_3$.

An important characteristic of this model is the "band filling" which physically corresponds to the ratio of the number of electrons divided by the number of lattice sites. For example, one electron per site corresponds to a half-filled band.

Experimentally, the band filling can be varied by doping of a specimen by electron donors or acceptors.

Using this Hamiltonian and the so called "memory function" method [11], a full analytical calculation of the conductivity of Q1D organic conductors was performed [12]. The main equations of the "memory function" approach are:

$$\chi_{AB} = -i\int_0^\infty \exp(izt)\langle[A(t),B(0)]\rangle dt \quad (2)$$

$$\sigma(\omega) = i\frac{\omega_P^2}{4\pi z}[1 - \frac{\chi(z)}{\chi_0}] \quad (3)$$

where $A = B = [j,H]$ and $j$ denotes the current operator:

$$j = -it\sum_{l,\sigma}\left(c_{l,\sigma}^+ c_{l+1,\sigma} - c_{l+1,\sigma}^+ c_{l,\sigma}\right) \quad (4)$$

Full details of the calculation, in which the Fermi liquid model was used, are available in the literature [12]. As an illustration, note that the sum invoked in the calculation of the susceptibility has about 2700 terms! Obviously, such a result can not be applied. Therefore the main problem of this calculation was how to introduce realistic approximations in it, so as to make it applicable, and to take into account the main terms in various sums in it.

No attempt was made to fit any particular set of experimental data. However, choosing various parameters in analogy with their values for high $T_c$ superconductors and then determining the temperature dependence of the conductivity, gave results in good agreement with experiment. A new expression for the chemical potential of the electron gas on a one dimensional lattice was obtained [13]. It was shown there that the chemical potential of the electron gas on a 1D lattice can be approximated as:

$$\mu = \frac{(\beta t)^6 (ns-1)|t|}{1.1029 + 0.1694(\beta t)^2 + 0.0654(\beta t)^4} \quad (5)$$

where $\beta$ denotes the inverse temperature, $n$ is the band filling, $t$ the hopping and $s$ the lattice constant. It follows from eq.(5) that the chemical potential depends on the band filling, which means on the doping.

A very important and interesting problem in studies of Q1D organic metals is the choice of the theoretical model which can be applied to the electron gas: Luttinger or Fermi liquid? Statistical mechanics claims that the standard Fermi liquid model fails in one dimension, and that in such systems the Luttinger liquid model should be applied [14]. However, Bechgaard salts are in reality Q1D, although they are mathematically treated as 1D. The opinion of the present author, based on personal experience and the available literature, is that Bechgaard salts in different situations can be described by both models.
However, the problem is open for further study, because there are numerous proofs and counter-proofs of both opinions.

What about the phase diagram of the Bechgaard salts? In theory, it can be determined by starting from eq. (1) and calculating the thermodynamic potentials and their singularities by the principles of statistical mechanics. Trying to perform a calculation in that way, one would easily fall into the usual "trap" of the Hubbard model: it only seems to be simple, and it is far from being so. A recent example of a thermodynamic study of the Hubbard model in 2D is given in [15]. Note that all the calculations in that paper were performed numerically and not analytically, which is a good illustration of the complexity of the problem. An interesting result of this paper is that the entropy depends on the band filling, and reaches a maximum near half filling, independently of the ratio $U/t$. It would be useful to perform a calculation of the conductivity, and check its behaviour as a function of the filling.

An even more complicated example of a determination of the phase diagram of the Hubbard model has appeared a short time ago [16].

The aim of this work was to determine the phase diagram of a one-dimensional model of interacting electrons, within the so called extended Hubbard model. This model is drastically more complex than the Hubbard model described by our eq.(1), as it is specified by three different interaction parameters: the on-site interaction U, nearest-neighbor V and pair-hopping interaction W. The calculation was performed in the continuum limit approach of field theory. In experimentally relevant terms, this means that this calculation could not take in to account effects due to the existence of the crystal lattice. This is a drawback, but at the same time facilitates the calculations. It was obtained that this model has a rich phase diagram, which contains a superconducting phase, a metallic phase and four different insulating phases. The applicability of the results of this work to the organic conductors has not been discussed so far, but it should be noted that the phase diagrams presented in this paper bear strong resemblance to experimental results on the organic conductors.

## 4. Two-dimensional systems

A short mention of the Q2D organic conductors is necessary, both because of their scientific interest and in order to render this review as complete as possible.

These salts are based on bisethylendithio-tetrathia-fulvalene, abbreviated as BEDT-TTF, or ET. Their chemical composition can be represented by $(ET)_m X_n$, where $X$ denotes an anion. However, in most cases $m:n = 2:1$. Synthesis of $ET$ salts was achieved by attempting to increase experimentally the dimensionality of Q1D salts [17]. The actual dimensionality of these salts is deter-mined by a competition between the non-planar structure of $ET$, and the large thermal vibrations of - $CH_2$ - [1]. The phase diagrams of Q2D salts are most often studied by conductivity measurements. It turns out that a wide variety of electrical properties exist, with materials ranging from insulators to superconductors.
Apart from this, dimensional crossover is also possible, depending on the ratio of the strengths of face to face and side by side interactions of $ET$ molecules [1]. Parameters which characterize the band structure of the ET salts have been deteremined from their reflectance spectra. Theoretical analysis was performed within the Drude model, but in which it was assumed that the plasma frequency and the relaxation rate were anisotropic. One of the results was the determination of the effective masses [1].

With all the avaliable experimental data on the conductivity of Q2D salts, it is tempting to undertake a theoretical study of the problem, taking into account the influence of as many as possible real material parameters. It is particularly interesting to try to take into account in this calculation the existence of a crystal lattice, as this gives the possibility to analyze the influence of high external pressure on the conductivity.

It can be expected that doping will influence the chemical potential,so the first step was to find a link between the chemical potential, filling factor and other measurable parameters of the system . In this calculation the electron gas was modelled as a Fermi liquid.

A general expression which links the chemical potential of the electron gas on a two dimensional lattice with various parameters of the system was recently obtained [18]. One of the results of this work is that the so-called Lieb-Wu theorem, well known in statistical physics, is in this case applicable only under certain conditions. In the continuation, a calculation of the conductivity will be undertaken, using the same "memory function" method as in the case of Q1D salts. One of the conclusions of this future work will be an independent argument in the discussions concerning modellisation of the electron gas in low dimensional systems.

## 5. Some general comments

Both Q1D and Q2D organic conductors are presently considered as examples of low dimensional correlated electron systems. They are scientifically interesting in their own rights, but it is also hoped that results and insights gained in studies of these systems will be a useful step towards the comprehension of high $T_c$ superconductors. The phas diagrams of high $T_c$ superconductors and the low dimensional organic conductors are known to show many similarities [19]. In spite of world-wide efforts, the problem of the origin of high temperature superconductivity still evades a definitive solution.

However, there exists and intermediary step: conductors based on fullerene. Fullerene ( $C_{60}$ ) has been discovered in 1985., as a result of work in chemistry,solid state physics and astrophysics. Although fullerene is a 3D system, and apparently does not belong to the field of low dimensional organic conductors, many of its characteristics are similar to those of the organics [1] .

Pure crystal fullerene is an insulator at room temperature. However, compounds (called fullerides) such as $Cs_3 C_{60}$ are superconducting under high external pressure, and critical temperatures of the order of 40 K have been detected for some of them. This value is smaller than those measured for the high temperature superconductors, but higher than those for the organic conductors. In a sense, the chain of rising values of $T_c$ contains three "steps": the organics, fullerides and high $T_c$ . The hope is that research work on materials which belong to all three steps of this chain will contribute to the full understanding of the superconductors.

## Acknowledgements
This work is a part of project 1231 financed by the MNTRS in Beograd. I am grateful to D. Jérome and the late H. J. Schulz for generating my interest in organic conductors, and to the editor for helpful comments.